\documentclass[conference]{IEEEtran}
\IEEEoverridecommandlockouts

\usepackage{cite}
\usepackage{amsmath,amssymb,amsfonts}
\usepackage{algorithmic}
\usepackage{graphicx}
\usepackage{textcomp}
\usepackage{xcolor}
\usepackage{algorithm}
\usepackage{booktabs}
\usepackage{multirow}
\usepackage{makecell}
\usepackage{orcidlink}
\def\BibTeX{{\rm B\kern-.05em{\sc i\kern-.025em b}\kern-.08em
    T\kern-.1667em\lower.7ex\hbox{E}\kern-.125emX}}
\begin{document}

\title{Dual-Channel Feature Fusion for Joint Prediction in Dynamic Signed Weighted Networks}

\author{Gaoxin~Zhang\orcidlink{0009-0007-9670-4975},
      Ruixing~Ren\orcidlink{0009-0000-1183-9512},
      Junhui~Zhao\orcidlink{0000-0001-5958-6622},~\IEEEmembership{Senior Member,~IEEE,}
      and Xiaoke~Sun
      
  \thanks{This work was supported in part by National Natural Science Foundation of China under Grant U25B2007, in part by the Fundamental Research Funds for the central Universities under Grant 2025JBZX060, in part by National Engineering Research Center of System Technology for High-Speed Railway and Urban Rail Transit under Grant 2024YJ255. (Corresponding author: Junhui Zhao.)}
  \thanks{Gaoxin Zhang, Ruixing Ren, Junhui Zhao are with the School of Electronic and Information Engineering, Beijing Jiaotong University, Beijing 100044, China. (e-mail: g72815767@hotmail.com; renruixing0604@163.com; junhuizhao@hotmail.com)}
  \thanks{Xiaoke Sun, the National Compter Network Emergency Response TechnicalTeam/Coordination Center of China (CNCERT/CC), Beijing 100029, China (e-mail: xiaokesun@cert.org.cn).}}

\maketitle

\begin{abstract}
Link prediction is central to unraveling social network evolution and node relationships, as well as understanding the characteristic mechanisms of complex networks. Currently, research on link prediction for complex dynamic networks integrating temporal evolution, relational polarity and edge weight information remains significantly underexplored, failing to meet practical demands. For dynamic signed-weighted networks, this paper proposes a tripartite joint prediction framework for unified forecasting of links, signs and weights. First, the dynamic network is decomposed into temporal snapshots, and node semantic embeddings are generated via sign-aware weighted random walks. We then design multi-hop structural balance and temporal difference features to capture the structural characteristics and dynamic evolution laws of the network, respectively. The model adopts a dual-channel feature decoupling mechanism: node semantic embeddings are used for link existence prediction, while relational sign features are fed into a Transformer encoder to model temporal dependencies. Finally, prediction results are output synergistically through a multi-task unit. Simulation experiments demonstrate that, compared with baseline methods, the proposed framework achieves an average 2\%–4\% improvement in the performance of link existence and relational sign prediction, and a significant 40\%–50\% reduction in edge weight prediction error.
\end{abstract}


\section{Introduction}

With the rapid development of online media, social networks have become increasingly complex and diverse. In this context, link prediction, a core problem in network science, has become a key technique for understanding and mining network structure and evolution \cite{1-3}. By analyzing existing network topologies, it aims to predict potential future connections or identify unobserved edges. This task is closely linked to applications such as friend recommendation, information diffusion prediction, and community detection, while its advances drive theoretical and methodological innovations in complex networks \cite{1-13}.

The development of link prediction methods is closely tied to the evolution of network types. Early research focused on static, homogeneous networks, aiming to mine potential links from fixed topologies. For example, Sahu et al.\cite{1-1} proposed an efficient algorithm for hub nodes in large networks, while Su et al.\cite{1-2} used diffusion models to simulate priors and mitigate noise. However, real-world networks exhibit high complexity in structure, attributes, and dynamics. Static models fail to capture continuous evolution, edge weight heterogeneity, and relational polarity, thus limiting their practical applicability.

Accordingly, research has shifted toward dynamic networks to capture temporal structural evolution. Tao et al.\cite{1-26} quantified dynamic complexity via traffic density and communication range. Skarding et al.\cite{1-3} combined discrete and continuous dynamic graph neural networks to adapt to temporal changes. Zhou et al.\cite{1-4} integrated community and temporal attention to capture spatial and temporal variations. Choudhary et al.\cite{1-5} captured structural and temporal evolution to improve accuracy but relied mainly on topology, lacking contextual information. To address this, Liu et al.\cite{1-6} captured contextual features from graph patterns and link sequences to enhance temporal link prediction. Similarly, in dynamic systems such as Internet of Vehicles, Zhao et al.\cite{1-22} used attention-based context modeling for high-precision prediction of time-varying states. Jiao et al.\cite{1-7} achieved unsupervised future link prediction by automatically capturing nonlinear structural and temporal features. However, these methods focus primarily on temporal topological changes and often overlook critical information such as link strength and relational properties.

In fact, links in real networks often possess both strength and polarity. In weighted networks, Zhao et al.\cite{1-8} proposed a path force-based link prediction algorithm. Ding et al.\cite{1-9} captured local weighted patterns and global dependencies, proposing a contrastive learning method. Liu et al.\cite{1-10} reduced network size significantly while preserving link information via node merging, self-loops, and edge fusion. When weights and dynamics coexist, link prediction faces higher-dimensional challenges, such as edge weight variation and node non-stationarity. To this end, Zhao et al.\cite{1-25} demonstrated the critical role of explicitly modeling periodic temporal diffusion in capturing complex evolution. Mei et al.\cite{1-12} captured structural and temporal features and introduced an alignment module to handle non-fixed node sets for dynamic environments. Lei et al.\cite{1-13} used Generative Adversarial Networks to enhance the generation of subsequent weighted network snapshots, leveraging dynamics, topology, and evolution.

With the deep study of dynamic weighted networks, researchers have shifted focus to signed networks. Chen et al.\cite{1-14} considered path lengths and signed link counts to better characterize node correlations, but the model only applies to static undirected signed networks, lacking directed analysis. To address this, He et al.\cite{1-15} proposed a dual-stream signed directed GCN to extract transitive and directional signed features, but it is also static and ignores temporal evolution. Luo et al.\cite{1-16} mapped embeddings to hyperbolic space to model high-order relations and hierarchies, enhancing node representations. Xu et al.\cite{1-17} encoded multi-order signed proximities via low-rank matrix approximation of node representations. Zhang et al.\cite{1-18} predicted relation types by incorporating preference proximity and multi-dimensional negative signed features. However, all these methods rely on static assumptions and fail to model the temporal evolution of signed networks.

To fill this gap, Dang et al.\cite{1-19} treated graph structures as time-series data and captured dynamic patterns via Recurrent Neural Networks (RNNs). Sharma et al.\cite{1-20} combined balance theory with temporal signed interactions, using memory modules and balanced aggregation to predict signed link evolution. Notably, real networks often exhibit dynamics, signedness, and weights simultaneously. For example, in Internet of Vehicles, networks handle highly dynamic topologies, heterogeneous communication strengths, and time-varying interaction patterns \cite{1-23, 1-24}. Sun et al.\cite{1-21} designed a dynamic signed collaboration unit for dynamic signed weighted networks to simulate the propagation and aggregation of signed semantics. Although effective for edge weight prediction, it does not explicitly target link existence, limiting its applicability to full link prediction tasks.

In summary, current research has two limitations: first, despite extensive studies on dynamic, signed, and weighted networks, research on their joint modeling remains scarce; second, existing methods for dynamic signed weighted networks suffer from limited generalization and cannot be extended to non-signed networks, restricting applicability.

To address these issues, this paper proposes an Link-Sign-Weight Joint Prediction (LSWJP) framework. It constructs node representations using edge information for semantic embeddings, multi-hop structural balance features, and temporal difference features. It uses dual-channel spatial-temporal encoding to decouple node features while capturing temporal dependencies, and finally performs joint prediction of links, signs, and weights. The main contributions of this paper are as follows:

\begin{itemize}
    \item A node representation method is proposed that fuses sign-aware weighted random walks, multi-hop structural balance features, and temporal difference features, which effectively captures semantic and evolutionary information in dynamic signed weighted networks.
    
    \item A dual-channel spatial-temporal encoding architecture is constructed to decouple semantic and structural signals. Transformers and temporal attention pooling are utilized to model dynamic dependencies, enabling joint prediction of links, sign polarities, and edge weights.
    
    \item Comprehensive evaluations on multiple real-world datasets are conducted. It is demonstrated that LSWJP reduces error by approximately 40\% for weight prediction while maintaining stable overall performance on link and sign prediction.
\end{itemize}

The rest of this paper is organized as follows: Section \ref{sec:problem} defines the joint prediction problem and evaluation metrics. Section \ref{sec:model} presents the LSWJP model, including network discretization, sign-aware node representation, multi-hop structural balance and temporal difference features, and the dual-channel spatial-temporal encoder-decoder architecture. Section \ref{sec:exp} details the experimental setup and results. Section \ref{sec:conclusion} concludes the paper.

\section{Basic Definitions}\label{sec:problem}
\subsection{Problem Statement}

A dynamic signed weighted network for link prediction is defined as $G=(V, E, W, S)$, where the four components represent the sets of nodes, edges, edge weights, and edge signs, respectively. For any two nodes $v_i, v_j \in V$, each edge $e_{v_i v_j} \in E$ is associated with a weight $w_{v_i v_j} \in W$ and a corresponding sign $s_{v_i v_j} \in S = \left\{ +, - \right\}$.

The link prediction task aims to predict whether a future connection will appear between currently unconnected node pairs based on the given network. This task is typically formalized as a binary classification problem: the model takes a node pair $(v_i, v_j)$ as input and outputs a score $\text{score}(v_i, v_j)$ or a probability $p(v_i v_j)$ indicating the likelihood of a link existing. The core objective is to assign scores to all candidate node pairs and rank them, with the highest-ranked pairs considered most likely to form links. During model training, observed connections are usually treated as positive samples, while unconnected node pairs are treated as negative samples.

The edge weight prediction task, targeting weighted networks, aims to predict the weight value of an existing edge or a future potential edge. This task can be formalized as a regression problem: given a node pair $(v_i, v_j)$, the model outputs a continuous estimate $w_{v_i v_j}$ of the edge weight.

The edge sign prediction task, designed for signed networks, predicts the sign of a future link between unconnected node pairs or infers the sign of an existing link when it is unknown. This task is also formalized as a binary classification problem: given a node pair $(v_i, v_j)$, the model outputs a sign $s_{v_i v_j} \in \left\{ +, - \right\}$, or the probability $p(s_{v_i v_j} = +)$ that the sign is positive.

\subsection{Evaluation Metrics}
\noindent\textit{(1) Binary Classification Metrics}

The confusion matrix is the foundation for evaluating binary classification models, containing four elements: True Positive (TP), which refers to instances that are actually positive and correctly predicted as positive; False Negative (FN), which refers to instances that are actually positive but incorrectly predicted as negative; False Positive (FP), which refers to instances that are actually negative but incorrectly predicted as positive; and True Negative (TN), which refers to instances that are actually negative and correctly predicted as negative, as shown in Table \ref{tab:confusion_matrix}.

\begin{table}[t]
	\centering
	\caption{Confusion Matrix}
	\begin{tabular}{ccc}
		\toprule
		Predictive\textbackslash Actural Class & Positive & Negative \\
		\midrule
		Positive & TP & FP \\
		Negative & FN & TN \\
		\bottomrule
	\end{tabular}
	\label{tab:confusion_matrix}
\end{table}

Traditional binary classification metrics (such as Precision or F1 score) rely on a fixed classification threshold, with sensitivity to the class distribution in the test set. The essence of link prediction and sign prediction is to provide a global ranking of the likelihood of connections between node pairs. Therefore, this paper adopts the Area Under the Curve (AUC) as the comprehensive evaluation metric for binary classification tasks. AUC is defined as the area under the Receiver Operating Characteristic curve, which plots the True Positive Rate against the False Positive Rate. AUC measures the overall ability of a model to rank positive instances ahead of negative ones across all classification thresholds, making it insensitive to class distribution and independent of threshold selection. It provides a comprehensive assessment of the model's ranking performance, thus being chosen as the core evaluation metric for link prediction and sign prediction.

\noindent\textit{(2) Regression Metrics}

For the edge weight prediction task, Mean Absolute Error (MAE) and Root Mean Squared Error (RMSE) are used to measure the deviation between predicted and true weight values. They are defined as follows:
\begin{equation}
	\text{MAE}=\frac{1}{n} \sum_{i=1}^{n}|y_i-\hat{y}_i|
\end{equation}
\begin{equation}
	\text{RMSE}=\sqrt{\frac{1}{n}\sum_{i=1}^{n}(y_i-\hat{y}_i)^2}
\end{equation}
where $y_i$ and $\hat{y}_i$ represent the true value and the model-predicted value of the $i$-th sample, respectively. MAE reflects the absolute average level of prediction errors, with lower sensitivity to outliers, while RMSE amplifies the impact of larger errors through the squared term, revealing more about the fluctuation of prediction errors.

\begin{figure*}[t]
	\centerline{\includegraphics[width=6.4in,keepaspectratio]{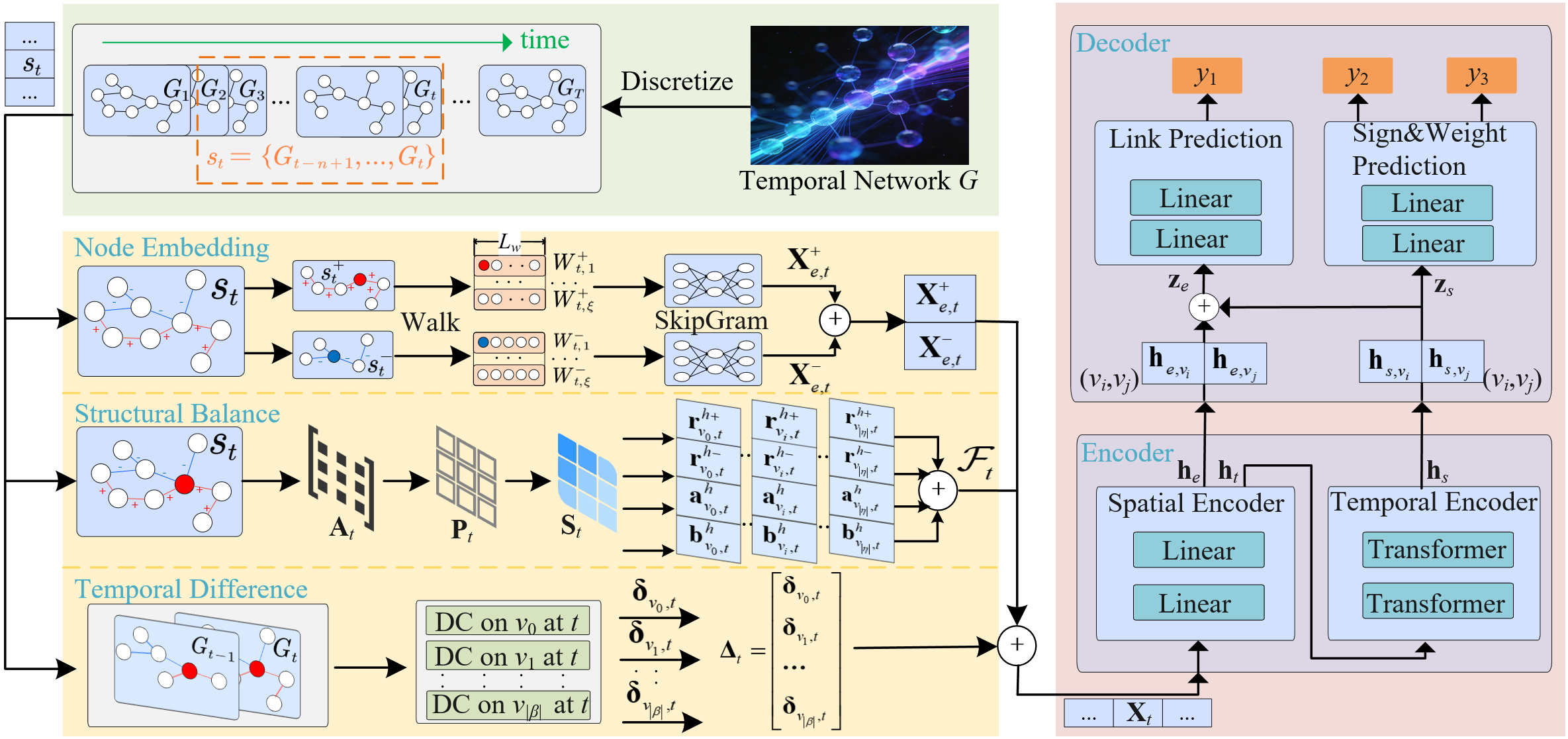}}
	\caption{Overall architecture diagram of the proposed LSWJP}
	\label{LSWJP}
\end{figure*}

\section{Model Design}\label{sec:model}
This section introduces the architecture of the proposed LSWJP framework, whose overall structure is shown in Fig. \ref{LSWJP}.

\subsection{Dynamic Network Discretization}
To address continuous-time dynamic networks, this paper discretizes the time axis into a sequence of ordered static snapshots $G_1, G_2, \dots, G_T$. Specifically, based on edge timestamps, the entire observation period is divided into $T$ equal-width time steps using a time binning strategy. Edges within each time step are aggregated to construct the corresponding graph snapshot $G_t$, to facilitating subsequent feature extraction and temporal encoding.

Predicting $G_{t+1}$ solely from the current snapshot $G_t$ makes it difficult for the model to capture temporal evolution trends, while using all historical information may introduce redundancy or noise and increase computational cost. To address this, the model adopts a sliding historical window to construct the input time series: for the target time step $t+1$, the most recent $n$ snapshots are selected to form the input sequence $s_t = \{G_{t-n+1}, \dots, G_t\}$. A spatiotemporal encoder jointly models temporal dependencies and structural dynamics to perform multi-task prediction of link existence, sign, and weight for $G_{t+1}$.

\subsection{Node Feature Construction}
The node features constructed in this paper consist of the following three components.

\noindent\textit{(1) Positive and Negative Node Embeddings.}

Graph embedding maps sparse graph structures into low-dimensional dense vectors, preserving topological and semantic relationships among nodes while maintaining computational efficiency. Considering that positive edges (indicating friendship, cooperation, etc.) and negative edges (indicating hostility, conflict, etc.) in signed networks exhibit semantic opposition, this paper proposes a sign-aware weighted random walk embedding method.

Given the signed network sequence $s_t$ within the historical window, we first decompose it by edge sign into a positive subgraph $s_t^+$ and a negative subgraph $s_t^-$. On each subgraph, we perform weight-aware random walks, where the transition probability is defined as:
\begin{equation}
    p_{v_i v_j} = \frac{|w_{v_i v_j}|}{\sum_{v_k \in \mathcal{N}_{v_i}} |w_{v_i v_k}|}
\end{equation}
where $w_{v_i v_j}$ is the weight of edge $(v_i, v_j)$, and $\mathcal{N}_{v_i}$ denotes the neighbor set of node $v_i$ in the current subgraph.

Based on this transition probability, we generate random walks of length $L_w$ starting from nodes in subgraph $s^\gamma$, $\gamma \in \{+, -\}$. The $\xi$-th walk sequence is expressed as:
\begin{equation}
	W^\gamma_{t,\xi} = \text{RandomWalk}(v_0, L_w, s^\gamma_t, p_{v_i v_j}),
\end{equation}
where $v_0$ is the starting node, and the generated sequence takes the form $W^\gamma_{t,\xi} = \{v_0, v_1, \dots, v_{L_w-1}\}$. The collection of all walk sequences on subgraph $s_t^+$ is denoted as $\mathcal{W}^+_t = \{W^+_{t,\xi} \mid \xi = 1, 2, \dots, \Xi\}$, where $\Xi$ is the number of walks. Similarly, the collection on subgraph $s_t^-$ is denoted as $\mathcal{W}^-_t$.

Subsequently, to learn co-occurrence patterns among nodes in these sequences and transform them into low-dimensional vectors that capture node associations in a machine-readable form, we feed the two sets of walk sequences into Skip-Gram models for training:
\begin{equation}
	\mathbf{X}^+_{e,t} = \text{SkipGram}(\mathcal{W}^+_t),
\end{equation}
\begin{equation}
	\mathbf{X}^-_{e,t} = \text{SkipGram}(\mathcal{W}^-_t),
\end{equation}
where $\mathbf{X}^+_{e,t} \in \mathbb{R}^{|\mathcal{V}| \times d}$ and $\mathbf{X}^-_{e,t} \in \mathbb{R}^{|\mathcal{V}| \times d}$ are the node embedding matrices obtained from the positive and negative subgraphs, respectively; $|\mathcal{V}|$ is the total number of nodes, and $d$ is the embedding dimension. Finally, we concatenate the positive and negative embeddings to obtain the final node representation at the $t$-th time window:
\begin{equation}
	\mathbf{X}^\ast_{e,t} = \text{Concat}(\mathbf{X}^+_{e,t}, \mathbf{X}^-_{e,t})
\end{equation}
This embedding simultaneously preserves node characteristics in both positive and negative relational contexts, providing discriminative representations for downstream tasks.

\noindent\textit{(2) Multi-hop Structural Balance Features.}

This feature is designed based on the structural balance theory from sociology. The theory states that in a stable social system, relationships among any triplet of nodes tend to satisfy one of the following two balanced patterns: (1) all three edges are positive (e.g., “the friend of my friend is my friend”); or (2) two edges are negative and one is positive (e.g., “the enemy of my enemy is my friend”). Traditional methods typically quantify local balance by counting the number of balanced triangles in the network.

However, directly applying triangle-based statistics in dynamic network scenarios faces two major limitations: (1) each snapshot represents only an extremely short time slice of the dynamic network, resulting in sparse complete triangles, weak feature signals, and high variance; and (2) triangles are rigid local structures that fail to effectively leverage richer multi-hop path information in the network.

To address these issues, we propose a structural balance feature based on sign propagation over multi-hop paths. The core idea is to generalize balance assessment from local triangles to multi-hop paths: we allow positive or negative relations to propagate and accumulate along paths, and quantify a node’s local structural balance tendency by computing the total positive and negative influence it receives within a multi-hop neighborhood.

Specifically, we first construct the signed adjacency matrix $\mathbf{A}_t$ for the input sequence $s_t$. To prevent high-degree nodes from dominating the propagation process, we degree-normalize $\mathbf{A}_t$ to obtain the signed transition matrix:
\begin{equation}
	\mathbf{P}_t = \mathbf{D}_t^{-1} \mathbf{A}_t
\end{equation}
where $\mathbf{D}_t$ is the degree matrix of the network. Subsequently, we model the influence propagated between nodes via signed paths by accumulating multi-hop transition probabilities. The cumulative transition matrix within $h$ hops is defined as:
\begin{equation}
	\mathbf{S}_t^{h} = \sum\limits_{k=1}^{h} \mathbf{P}_t^{k}
\end{equation}
where $\mathbf{P}_t^k$ denotes the $k$-th power of the transition matrix, and $\mathbf{S}_t^h$ is the cumulative transition matrix. This approach fully exploits the network’s structural information, leading to more accurate influence estimation.

Based on $\mathbf{S}_t^h$, we further decompose the total positive and negative propagation received by each node:
\begin{equation}
	\mathbf{r}_{v_i,t}^{h+} = \sum \left| \mathbf{S}_{v_i,t}^{h} \right| + \mathbf{S}_{v_i,t}^{h}
\end{equation}
\begin{equation}
	\mathbf{r}_{v_i,t}^{h-} = \sum \left| \mathbf{S}_{v_i,t}^{h} \right| - \mathbf{S}_{v_i,t}^{h}
\end{equation}
where $\mathbf{S}_{v_i,t}^{h}$ denotes the row-wise sum of the matrix $\mathbf{S}_t^h$ corresponding to node $v_i$. We then construct two new features: the propagation balance coefficient, which reflects the overall positive/negative tendency of a node’s influence, and the total propagation strength, which measures its overall influence magnitude, defined as:
\begin{equation}
	\mathbf{b}_{v_i,t}^h = \frac{\mathbf{r}_{v_i,t}^{h+} - \mathbf{r}_{v_i,t}^{h-}}{\mathbf{r}_{v_i,t}^{h+} + \mathbf{r}_{v_i,t}^{h-}}
\end{equation}
\begin{equation}
	\mathbf{a}_{v_i,t}^h = \mathbf{r}_{v_i,t}^{h+} + \mathbf{r}_{v_i,t}^{h-}
\end{equation}
Finally, we concatenate the total positive propagation $\mathbf{r}_{v_i,t}^{+\left( h \right)}$, total negative propagation $\mathbf{r}_{v_i,t}^{-\left( h \right)}$, propagation balance coefficient $\mathbf{b}_{v_i,t}^h$, and total propagation strength $\mathbf{a}_{v_i,t}^h$ to form the multi-hop structural balance feature $\mathcal{F}_t$.

\noindent\textit{(3) Temporal Difference Features}

The topology and edge weights of dynamic networks continuously evolve over time. Such temporal variations not only reflect dynamic adjustments in relationships between nodes but also contain critical signals about the evolution of local structures. To explicitly model these dynamic patterns, this paper introduces temporal difference features that capture the changes in node attributes between consecutive time steps, enabling the model to perceive the evolutionary trends of positive/negative interaction intensities and edge weights.

This temporal difference feature consists of four core components: the differences in weighted positive/negative in-degrees and out-degrees of node $v_i$ between snapshots $G_t$ and $G_{t-1}$. The specific formulation is as follows:
\begin{equation}
\boldsymbol{\delta}_{v_i,t} = 
\begin{bmatrix}
    w_{\text{in},t}^{+} - w_{\text{in},t-1}^{+} \\
    w_{\text{out},t}^{+} - w_{\text{out},t-1}^{+} \\
    w_{\text{in},t}^{-} - w_{\text{in},t-1}^{-} \\
    w_{\text{out},t}^{-} - w_{\text{out},t-1}^{-}
\end{bmatrix}
\label{eq:delta_vector}
\end{equation}
where $w_{\text{in}}^{+}$, $w_{\text{out}}^{+}$, $w_{\text{in}}^{-}$, and $w_{\text{out}}^{-}$ denote the total weighted in-degree of positive edges, total weighted out-degree of positive edges, total weighted in-degree of negative edges, and total weighted out-degree of negative edges, respectively. Considering all nodes in the network, we stack the difference vectors of all nodes row-wise to form the network’s temporal difference feature matrix:
\begin{equation}
	\boldsymbol{\Delta}_t = \begin{bmatrix}
		\boldsymbol{\delta}_{v_0,t} \\
		\boldsymbol{\delta}_{v_1,t} \\
		\dots \\
		\boldsymbol{\delta}_{v_{|\beta|},t}
	\end{bmatrix}
\end{equation}
where $|\beta|$ is the number of nodes in the current snapshot $G_t$. Finally, we concatenate the node embedding representation $\mathbf{X}_{e,t}^\ast$, the multi-hop structural balance feature $\mathcal{F}_t$, and the temporal difference feature $\boldsymbol{\Delta}_t$ to obtain the final node representation $\mathbf{X}_t$, which is fed into the subsequent encoder–decoder module.

\subsection{Encoder}
The encoder in this paper consists of a dual-channel spatiotemporal encoding architecture, designed to separately model static semantic features and dynamic structural evolution features of nodes, thereby reducing task interference and enhancing representation generalization. Specifically, in the spatial encoder, we employ independent Multi-Layer Perceptrons (MLPs) to perform nonlinear transformation and feature decoupling: one branch processes the node embedding representation to output a semantic embedding vector $\mathbf{h}_e$, while the other branch processes the node structural feature representation to output a structural evolution vector $\mathbf{h}_t$, which is then fed into the temporal encoder for sequential modeling. For notational convenience, unless otherwise specified (notably for $\mathbf{h}_t$, which requires temporal operations), the time index $t$ is omitted from other vectors in the following descriptions.

The temporal encoding primarily adopts the standard Transformer encoder architecture, comprising Multi-Head Self-Attention, Feed-Forward Network, and Layer Normalization. This design enables the model to process the entire time sequence in parallel, significantly improving training efficiency. Moreover, by eliminating recurrent structures, it effectively avoids the gradient vanishing or explosion problems commonly encountered in traditional RNNs and Long Short-Term Memory networks, leading to more stable training.

The temporal encoder models the time series independently for each node to capture its unique evolutionary pattern. Specifically, the structural vectors within the historical window are concatenated into a sequence: $[\mathbf{h}_{t-n+1}, \mathbf{h}_{t-n+2}, \dots, \mathbf{h}_{t}]$. The Transformer encoder outputs context-aware features for each time step: $[\tilde{\mathbf{h}}_{t-n+1}, \tilde{\mathbf{h}}_{t-n+2}, \dots, \tilde{\mathbf{h}}_{t}]$. Subsequently, a learnable temporal attention pooling mechanism is employed to adaptively aggregate features along the time dimension. This mechanism computes an importance score for each time step $\tau$ via a two-layer perceptron:
\begin{equation}
	e_{\tau} = \mathbf{u}^\intercal \text{tanh}(\mathbf{U} \tilde{\mathbf{h}}_\tau + \mathbf{b})
\end{equation}
where $\mathbf{u}$ and $\mathbf{U}$ are learnable parameter vector and matrix, respectively, and $\mathbf{b}$ is a bias term. Softmax normalization is applied to the scores across all time steps to obtain attention weights:
\begin{equation}
	\alpha_{\tau} = \frac{\text{exp}(e_{\tau})}{\sum\nolimits_{k=t-n+1}^{t} \text{exp}(e_{k})}
\end{equation}
These weights are then used to compute a weighted sum of the time-step features, yielding the final node structural vector:
\begin{equation}
	\mathbf{h}_s = \sum\limits_{\tau=t-n+1}^{t} \alpha_{\tau} \tilde{\mathbf{h}}_{\tau}
\end{equation}
This mechanism dynamically emphasizes historically discriminative time steps for the prediction task, enabling time-aware feature selection and effective spatiotemporal information fusion.

\subsection{Decoder}
Since link, sign, and weight prediction are edge-level tasks while the encoder outputs node-level representations, edge-level feature representations must be constructed before feeding $\mathbf{h}_e$ and $\mathbf{h}_s$ into the respective prediction heads. Specifically, for a candidate edge $(v_i, v_j)$, we concatenate the semantic embeddings and structural vectors of its two endpoint nodes to obtain $[\mathbf{h}_{e,v_i}, \mathbf{h}_{e,v_j}]$ and $[\mathbf{h}_{s,v_i}, \mathbf{h}_{s,v_j}]$, respectively. We further concatenate these two to form the edge representation $\mathbf{z}_e$ for link prediction, and use $[\mathbf{h}_{s,v_i}, \mathbf{h}_{s,v_j}]$ as the edge representation $\mathbf{z}_s$ for sign and weight prediction. This design preserves individual node semantics while explicitly modeling pairwise interaction patterns, supporting both functional decoupling and collaborative optimization across multiple tasks.

The decoder consists of two prediction units:

\noindent\textit{(1) Link Prediction Unit:}  
In real-world large-scale networks, existing edges typically constitute only a small fraction of all possible node pairs. Training on all non-edges would lead to severe class imbalance and excessive computational cost, causing the predictor to bias toward negative predictions. To address this, we adopt a dynamic negative sampling strategy: edges that appear at time step $t+1$ are treated as positive samples, and an equal number of negative samples are uniformly sampled from unconnected node pairs at the same time step, transforming link prediction into a balanced binary classification task. The link existence probability is computed as:
\begin{equation}
    \ell_1 = \mathbf{w}_1^\intercal \sigma(\mathbf{W}_1 \mathbf{z}_e + \mathbf{b}_1) + b_1^\prime
\end{equation}
where $\mathbf{w}_1$ and $\mathbf{W}_1$ are learnable parameter vector and matrix, $\mathbf{b}_1$ and $b_1^\prime$ are bias terms, and $\sigma(\cdot)$ denotes the ReLU activation function. The final output $y_1 \in \{0,1\}$, obtained via a Sigmoid function, indicates whether the edge exists ($y_1 = 1$ if the edge exists).

\noindent\textit{(2) Edge Weight and Sign Prediction Unit:}  
Given that a link is confirmed to exist, this unit further infers its numerical strength and relational polarity. It comprises two sub-tasks: sign classification and weight regression, producing the probability of a positive sign and the predicted weight value, respectively:
\begin{equation}
    \ell_2 = \mathbf{w}_2^\intercal \sigma(\mathbf{W}_2 \mathbf{z}_s + \mathbf{b}_2) + b_2^\prime
\end{equation}
\begin{equation}
    y_3 = \mathbf{w}_3^\intercal \sigma(\mathbf{W}_2 \mathbf{z}_s + \mathbf{b}_2) + b_3^\prime
\end{equation}
where $\mathbf{w}_2$, $\mathbf{w}_3$, and $\mathbf{W}_2$ are learnable parameter vectors and matrix, and $\mathbf{b}_2$, $b_2^\prime$, and $b_3^\prime$ are bias terms. The value $\ell_2$ is passed through a Sigmoid function to yield $y_2 \in \{0,1\}$, where $y_2 = 1$ indicates a positive sign. The output $y_3$ is a real-valued scalar representing the predicted edge weight.

\section{Experimental Analysis}\label{sec:exp}
\subsection{Experimental Datasets}

This paper selects four publicly available dynamic network datasets from the Stanford Large Network Dataset Collection website (https://snap.stanford.edu/data/), covering three representative scenarios: signed weighted, signed unweighted, and unsigned weighted networks. This selection enables a comprehensive evaluation of the model’s generalization capability across diverse network structures. The chosen datasets exhibit significant differences in node scale, edge density, and sign distribution, which helps validate the robustness of the proposed method. Detailed network statistics are shown in Table \ref{network}.

\begin{itemize}
    \item The \text{soc-sign-bitcoinalpha} and the \text{soc-sign-bitcoinotc} are dynamic signed weighted networks representing trust relationships among users trading with Bitcoin. Members rate each other on an integer scale from $-10$ (complete distrust) to $+10$ (complete trust), with a step size of 1.
    
    \item The \text{wiki-RfA} dataset is a voting network that records votes cast by users on Wikipedia Requests for Adminship (RfA). Each vote includes a rating ($-1$ for oppose, $0$ for neutral, $1$ for support), along with its timestamp. Based on this, we construct a dynamic signed unweighted network. Note that neutral votes ($0$) carry no relational meaning and thus do not correspond to edges; consequently, they are excluded during network construction.
    
    \item The \text{email-Eu-core-temporal} dataset is an email communication network that logs the timestamps of emails sent from source nodes to target nodes. In this work, we treat the number of interactions between two nodes as the edge weight, thereby constructing a dynamic unsigned weighted network.
\end{itemize}

\begin{table}[t]
	\centering
	\caption{Network Statistics}
	\begin{tabular}{ccccc}
		\toprule
		Network & Nodes & Edges & Positive Edges & Negative Edges \\
		\midrule
		bitcoinalpha & 3,783 & 24,168 & 22,650 & 1,518 \\
		bitcoinotc & 5,881 & 35,592 & 32,029 & 3,563 \\
		wiki-RfA & 10,279 & 177,259 & 137,966 & 39,293 \\
		email-Eu & 1,005 & 16,064 & 16,064 & 0 \\
		\bottomrule
	\end{tabular}
	\label{network}
\end{table}

\subsection{Baseline Methods}
This paper selects several recent link prediction models that have demonstrated strong performance on dynamic or signed networks as baselines for comparison:
\begin{itemize}
	\item \textbf{DyCATW} \cite{1-4}: This model assumes the network is an unsigned, unweighted dynamic graph. It jointly models spatial structural evolution and temporal dynamics through community-aware attention and temporal attention mechanisms.
	
	\item \textbf{SIHG} \cite{1-16}: Focused on static signed networks, this method learns node embeddings in hyperbolic space and enhances sign awareness by maximizing mutual information between edge polarities and embeddings.
	
	\item \textbf{SPMF} \cite{1-17}: This approach employs low-rank matrix factorization to learn node representations, capturing global balance structures in signed networks. However, it is designed for static scenarios and exhibits limited performance in dynamic link prediction tasks.
	
	\item \textbf{SEMBA} \cite{1-20}: Incorporating a memory-augmented mechanism and structural balance aggregation, SEMBA models high-order neighborhood evolution in dynamic signed networks. Its prediction target is limited to link existence only; it does not support continuous edge weight regression, and its generalization capability is constrained in highly sparse networks.
	
	\item \textbf{DynamiSE} \cite{1-21}: This method constructs a deep dynamic signed graph neural network using dynamic signed collaboration units to encode network dynamics and capture signed semantics among neighbors. However, it primarily focuses on node representation learning and weight prediction, neglecting the explicit goal of link prediction.
\end{itemize}

\subsection{Parameter Settings and Sensitivity Analysis}
To verify the effectiveness of the proposed model, we split the data into training and test sets in an 8:2 ratio. In terms of architecture, the encoder accepts inputs of dimension 136 and outputs 64-dimensional representations; the decoder receives 128-dimensional inputs. Both the MLPs and Transformer layers within the encoder–decoder modules operate in a 64-dimensional space. The temporal encoder consists of two Transformer encoder layers, each with 4 self-attention heads. The learning rate is set to 0.001 to facilitate convergence.

For node embedding, each node performs $\Xi = 5$ weighted random walks, each of length $L_w = 10$, yielding an embedding dimension of $d = 64$. Regarding snapshot construction, the entire time sequence is divided into $T = 100$ snapshots, and the historical window size is set to $n = 10$. The hop count for the multi-hop structural balance mechanism is set to $h = 2$ to balance local and higher-order neighborhood information.

\begin{figure}[t]
	\centerline{\includegraphics[width=3.1in,keepaspectratio]{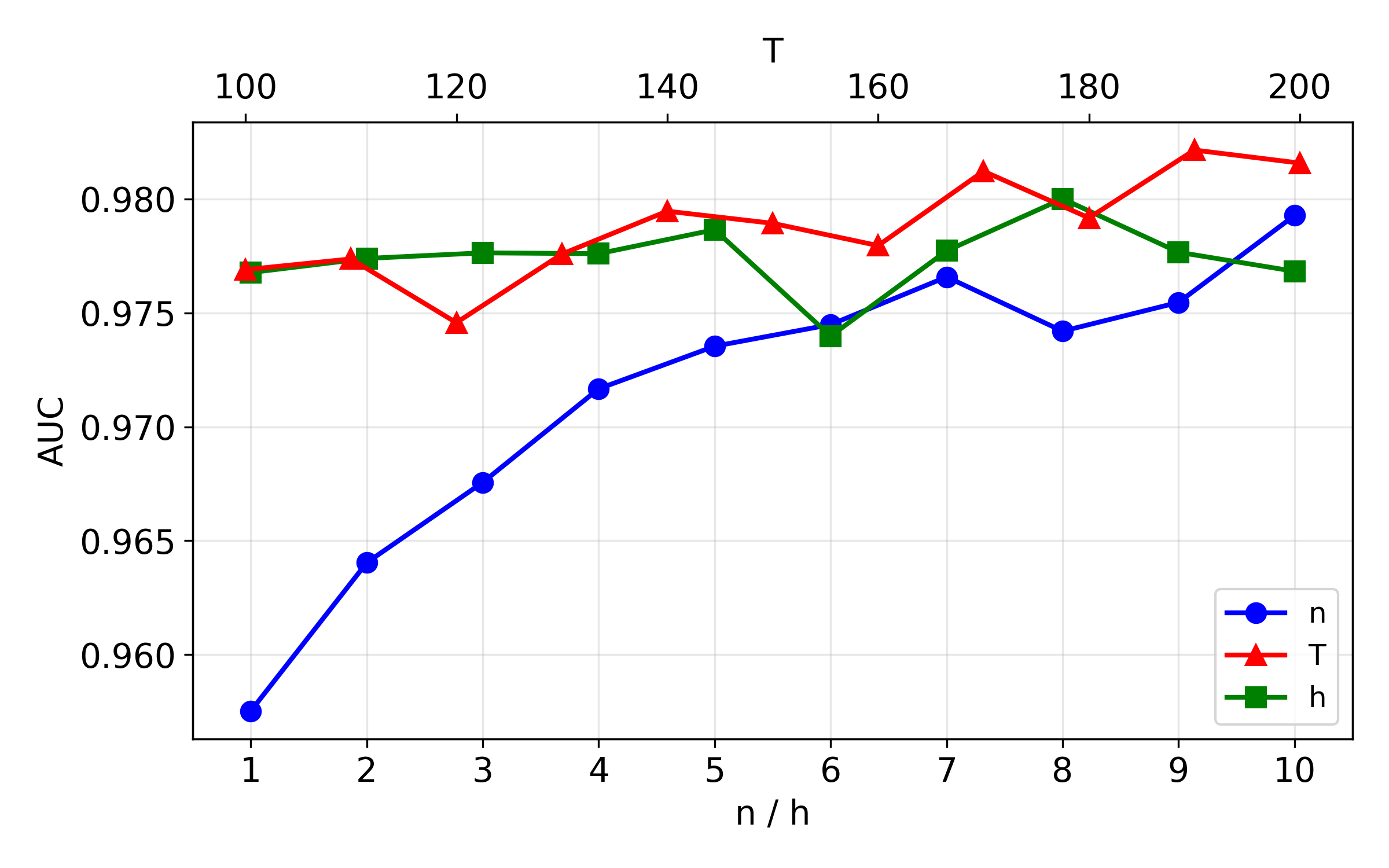}}
	\caption{Variation of link prediction AUC with hyperparameters}
	\label{auc_l_parameter}
\end{figure}
\begin{figure}[t]
	\centerline{\includegraphics[width=3.1in,keepaspectratio]{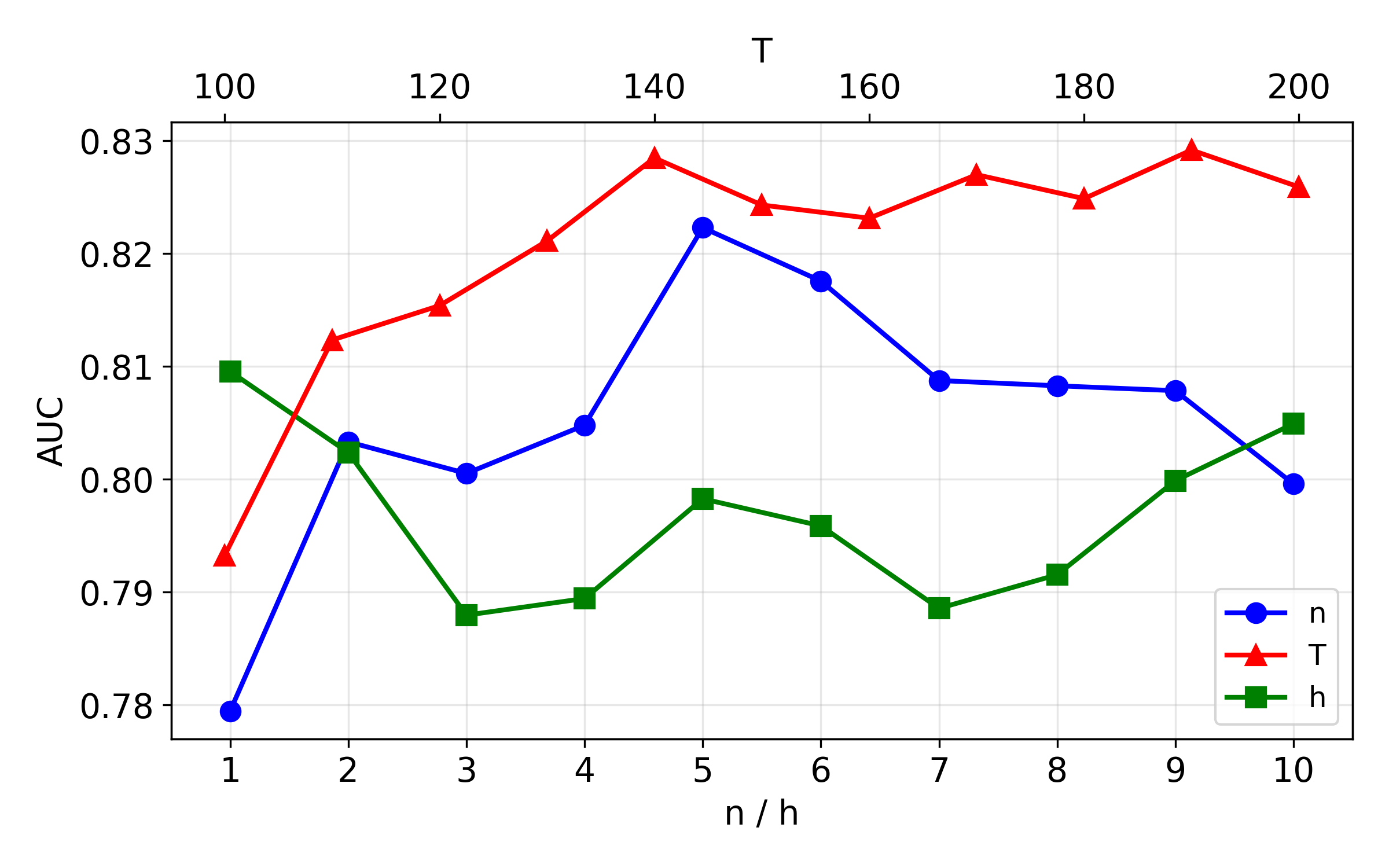}}
	\caption{Variation of sign prediction AUC with hyperparameters}
	\label{auc_s_parameter}
\end{figure}
\begin{figure}[t]
	\centerline{\includegraphics[width=3.1in,keepaspectratio]{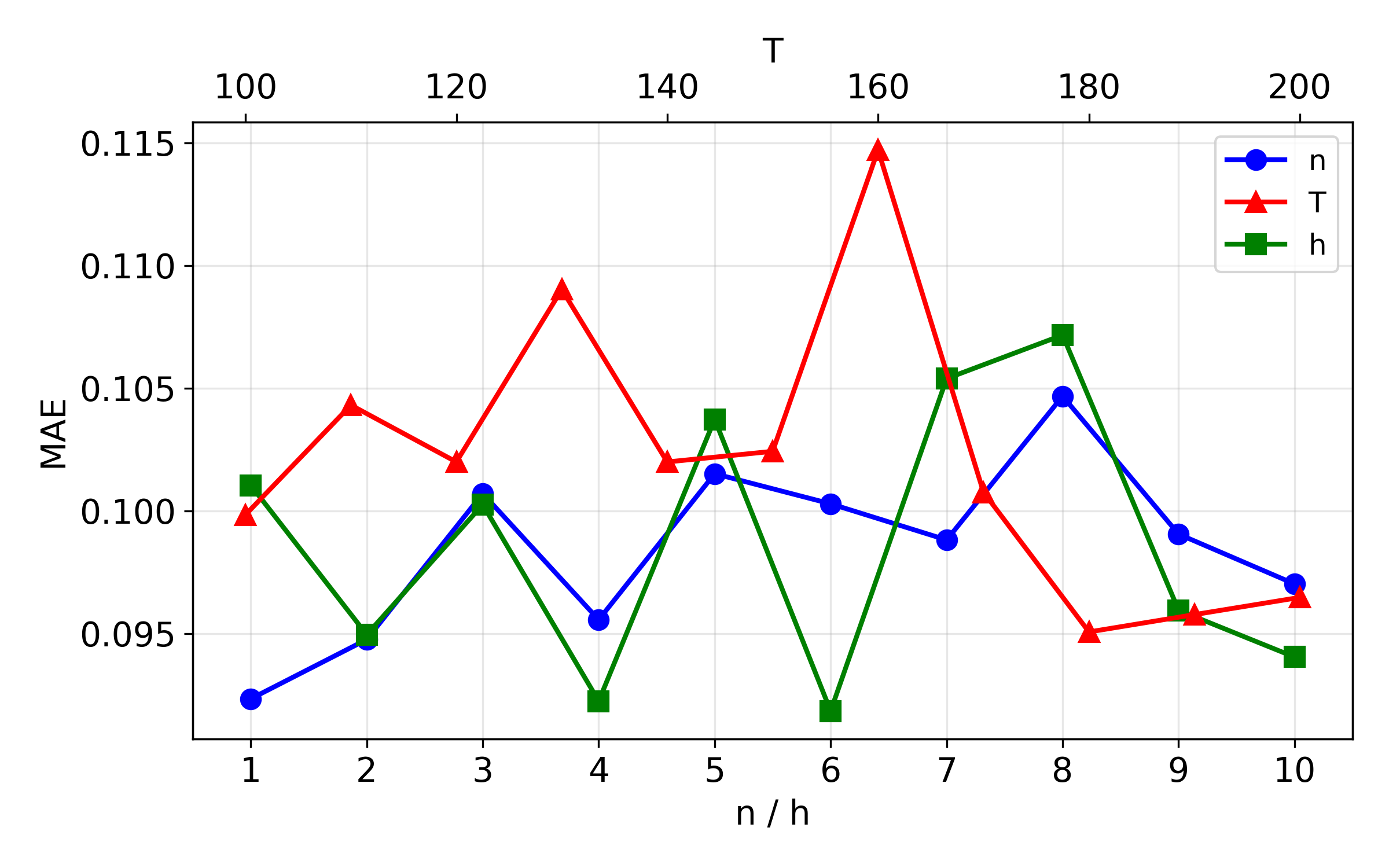}}
	\caption{Variation of weight prediction MAE with hyperparameters}
	\label{mae_parameter}
\end{figure}
\begin{figure}[t]
	\centerline{\includegraphics[width=3.1in,keepaspectratio]{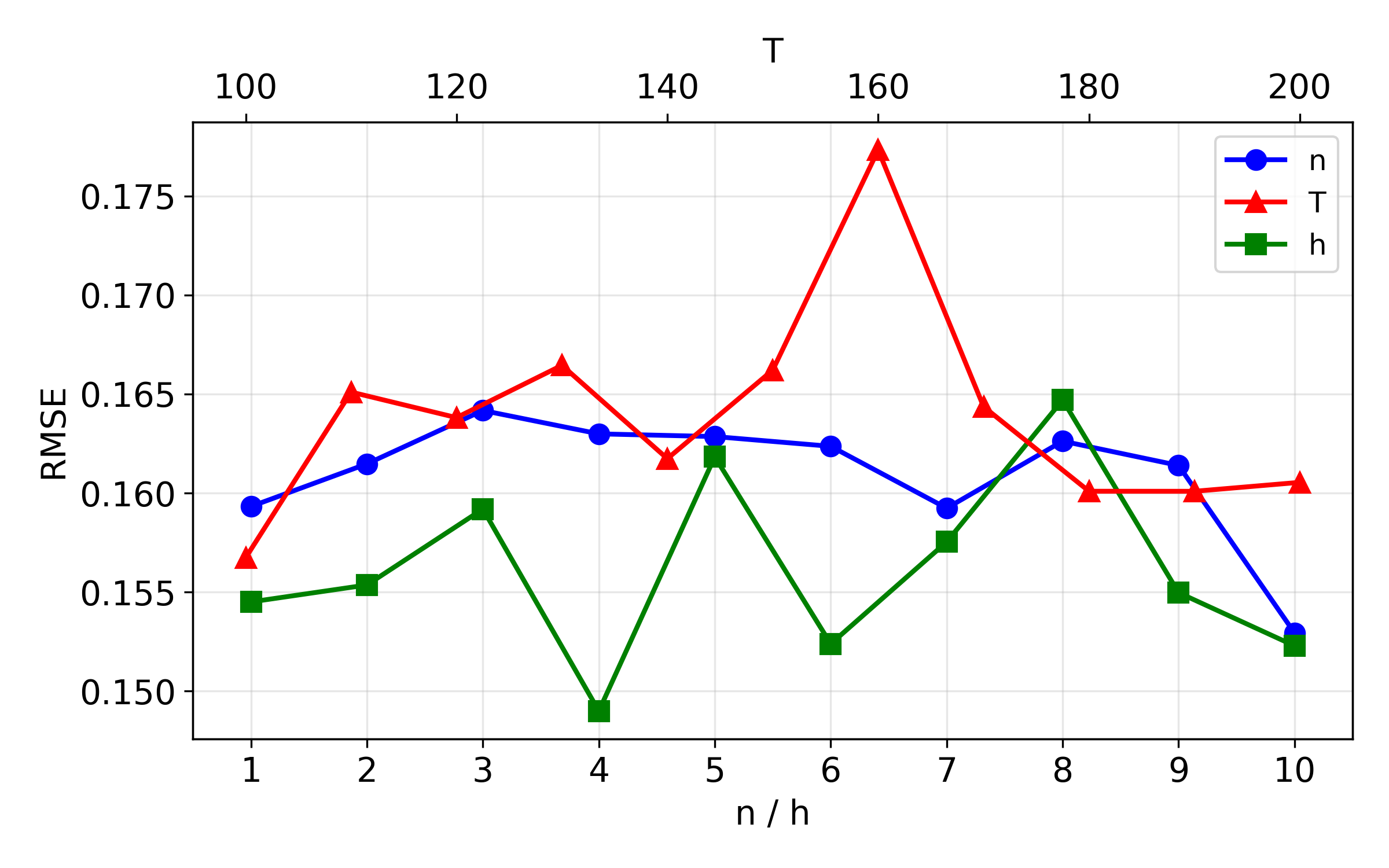}}
	\caption{Variation of weight prediction RMSE with hyperparameters}
	\label{rmse_parameter}
\end{figure}

The main computational cost of this method arises from the calculation of multi-hop structural balance features in the node feature construction stage. This operation relies on the power operation and accumulation of the normalized adjacency matrix $\textbf{P}$, with a time complexity of $O(h \cdot |\eta|^3)$ on the sequential subgraph $s_t$, where $|\eta|$ denotes the number of nodes in $s_t$. Since this operation must be executed sequentially on each sequential subgraph, the total complexity is $O((T-n) \cdot h\cdot |\eta|^3)$, where $T$ is the total number of snapshots and $n$ is the historical window size.

Given the above computational characteristics, this paper focuses on three key hyperparameters in the data preprocessing stage: the number of snapshots $T$, the historical window length $n$, and the hop count $h$. Among them, $T$ determines the information density per snapshot, $n$ affects the amount of historical information used for prediction, and $h$ influences the influence range of structural balance features. To systematically evaluate the impact of each parameter on performance, a parameter sensitivity analysis is conducted on the bitcoinalpha dataset. Specifically, the initial settings are $T=100$, $n=10$, and $h=2$. Then, the control variable method is adopted to adjust a single parameter at a time: $n$ is increased from 1 to 10, $h$ is increased from 1 to 10, and $T$ is increased from 100 to 200 in steps of 10, while other parameters remain unchanged.

\begin{figure}[t]
	\centerline{\includegraphics[width=3.1in,keepaspectratio]{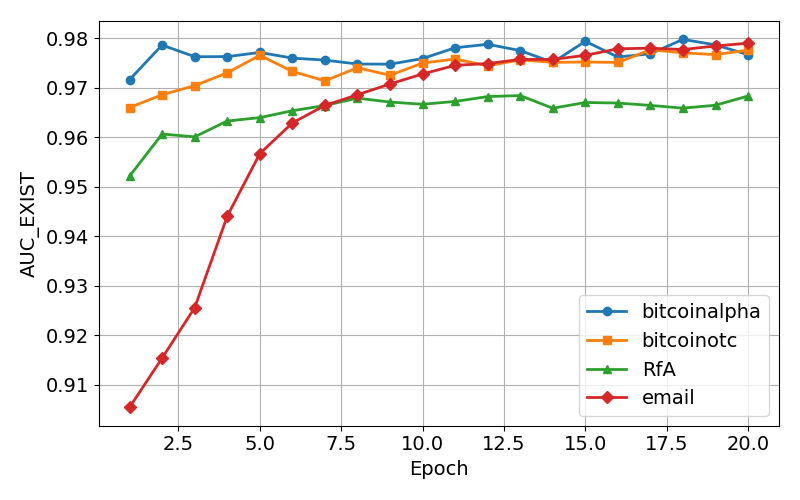}}
	\caption{Link prediction AUC across four datasets}
	\label{auc_l_comparison}
\end{figure}
\begin{figure}[t]
	\centerline{\includegraphics[width=3.1in,keepaspectratio]{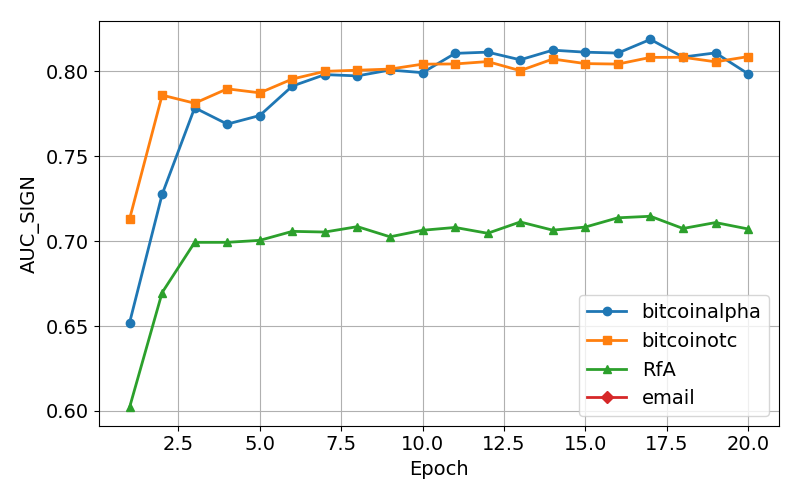}}
	\caption{Sign prediction AUC across four datasets}
	\label{auc_s_comparison}
\end{figure}
\begin{figure}[t]
	\centerline{\includegraphics[width=3.1in,keepaspectratio]{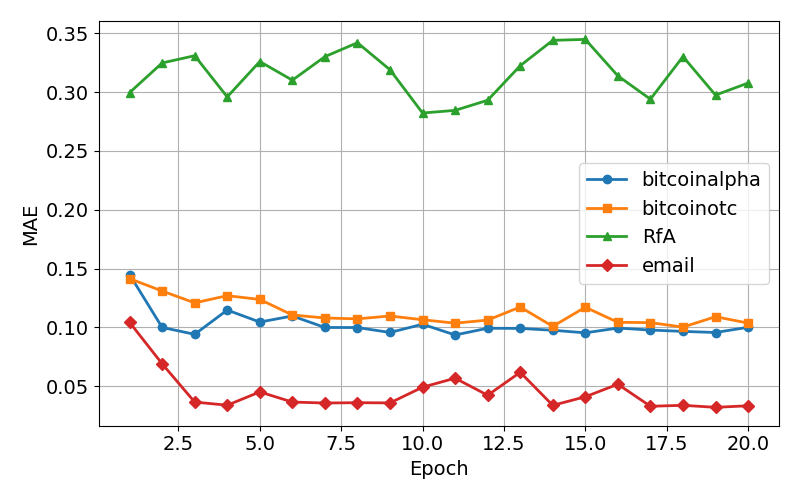}}
	\caption{Weight prediction MAE across four datasets}
	\label{mae_comparison}
\end{figure}
\begin{figure}[t]
	\centerline{\includegraphics[width=3.1in,keepaspectratio]{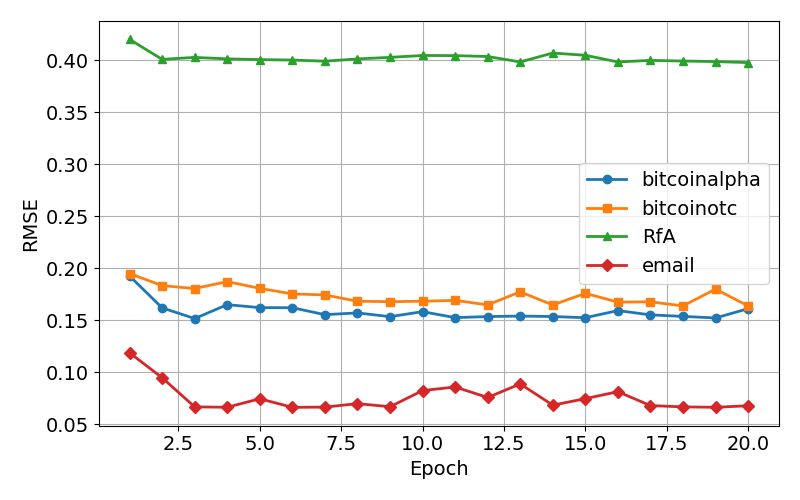}}
	\caption{Weight prediction RMSE across four datasets}
	\label{rmse_comparison}
\end{figure}



In Fig. \ref{auc_l_parameter}, the AUC of link prediction generally shows an upward trend as $n$ and $T$ increase. This is because link prediction mainly relies on node embeddings, finer-grained temporal segmentation and richer historical information help the generated node embeddings to more fully characterize node linkage information, thereby improving the quality of node embeddings. In contrast, the AUC is insensitive to changes in the hop count $h$, indicating that link existence is mainly dominated by local neighborhood structure and semantic embeddings—while high-order structural balance information has a supplementary role, its impact is limited.

In Fig. \ref{auc_s_parameter}, the AUC of sign prediction first increases and then decreases as $n$ increases, peaking when $n=5$, and gradually improves and stabilizes as $T$ increases. This suggests that a moderate historical window length is beneficial for capturing patterns in time series, while an excessively long window may introduce excessive noise. Meanwhile, the AUC first decreases and then increases as $h$ increases, indicating that a small $h$ can focus on the node's adjacent range to effectively characterize local positive/negative propagation influence, while a large $h$ can integrate higher-order path information and focus more on the global influence range.

From Figs. \ref{mae_parameter} and Figs. \ref{rmse_parameter}, the trends of the two metrics for weight prediction with parameter changes are basically consistent, with the lowest errors occurring at $T=180$, $n=10$, and $h=4$. This implies that appropriate snapshot segmentation can effectively preserve short-term fluctuation information, while a moderately extended historical window helps reveal long-term trends, thereby improving the accuracy of weight prediction. At the same time, an appropriate structural balance hop count achieves a good balance between local details and global structure, avoiding influence dilution due to excessive diffusion and providing the most discriminative feature representation for the weight regression task.

\subsection{Experimental Results}

As shown in Fig. \ref{auc_l_comparison}, the proposed LSWJP achieves consistently strong performance across all datasets, with AUC values consistently around 0.97, demonstrating its effectiveness in capturing the dynamic evolution patterns of latent connections between nodes.

As illustrated in Fig. \ref{auc_s_comparison}, sign prediction performance varies across datasets. On \text{bitcoinalpha} and \text{bitcoinotc}, the AUC reaches approximately 0.80, but decreases to approximately 0.70 on the larger-scale \text{wiki-RfA} dataset. This degradation is primarily attributed to its significantly higher node count and more complex community structure, which increase the difficulty of learning sign patterns. Notably, since the \text{email-EU} network is unsigned, the model automatically disables the sign prediction task upon detecting an unsigned network, demonstrating its adaptability to diverse network types.

As shown in Figs. \ref{mae_comparison} and Figs. \ref{rmse_comparison}, LSWJP achieves MAE and RMSE values in the range of 0.1 to 0.4 on weighted datasets. Specifically, on \text{wiki-RfA}, where all edge weights have absolute value 1 (i.e., lacking meaningful weight variation), the model struggles to learn an effective weight mapping, resulting in relatively higher errors. In contrast, on the other three datasets that provide informative weight signals, both RMSE and MAE are reduced below 0.20, validating the effectiveness of our joint modeling of signs and weights.

To further verify the superiority of LSWJP, we compare the proposed method against baseline models on three tasks: link prediction, sign prediction, and weight prediction. The results are summarized in Table \ref{table_auc} and Table \ref{table_mae}. Note that since \text{email-EU} is an unsigned network and most baselines cannot handle such cases, it is excluded from the comparison tables.

\begin{table}[t]
\centering
\caption{Comparison of link and sign prediction performance}
\label{table_auc}
\setlength{\tabcolsep}{3pt}
\begin{tabular}{lcccccc}
\toprule
& \multicolumn{2}{c}{\footnotesize bitcoinalpha} 
& \multicolumn{2}{c}{\footnotesize bitcoinotc} 
& \multicolumn{2}{c}{\footnotesize wiki-RfA} \\
\cmidrule(lr){2-3} \cmidrule(lr){4-5} \cmidrule(lr){6-7}
Method & AUC(L) & AUC(S) & AUC(L) & AUC(S) & AUC(L) & AUC(S) \\
\midrule
DyCATW & 0.8973 & —      & 0.8945 & —      & 0.7750 & —      \\
SIHG   & 0.8847 & —      & 0.8851 & —      & 0.8421 & —      \\
SPMF   & 0.8910 & \textbf{0.8391} & 0.8653 & \textbf{0.8791} & 0.8930 & \textbf{0.8667} \\
SEMBA  & \textbf{0.9942} & 0.7002 & \textbf{0.9952} & 0.7902 & \textbf{0.9901} & 0.7105 \\
LSWJP  & 0.9769 & 0.8187 & 0.9768 & 0.8101 & 0.9678 & 0.7144 \\
\bottomrule
\end{tabular}
\end{table}

\begin{table}[t]
    \centering
	\caption{Comparison of weight prediction performance}
    \setlength{\tabcolsep}{5pt}
    \begin{tabular}{ccccccc}
        \toprule
        & \multicolumn{2}{c}{bitcoinalpha}  & \multicolumn{2}{c}{bitcoinotc}    & \multicolumn{2}{c}{wiki-RfA}      \\
        \cmidrule(lr){2-3} \cmidrule(lr){4-5} \cmidrule(lr){6-7}
        Method& MAE & RMSE & MAE & RMSE & MAE & RMSE  \\
        \midrule
        DynamiSE & 0.1548  & 0.2982  & 0.1782 & 0.3645 & 0.6550  & 0.8336  \\
        LSWJP  & \textbf{0.0977} & \textbf{0.1552} & \textbf{0.1000} & \textbf{0.1631} & \textbf{0.2949} & \textbf{0.4007} \\
        \bottomrule
    \end{tabular}
    \label{table_mae}
\end{table}

From Table \ref{table_auc}, although LSWJP is slightly inferior to SEMBA in link prediction and to SPMF in sign prediction, it significantly outperforms all other baselines and achieves a more balanced multi-task performance. Specifically, SPMF treats the dynamic network as a static signed graph during training, allowing it to leverage the global sign distribution as a strong prior—thus gaining an advantage in sign classification—but it ignores temporal evolution and local structural dynamics, limiting its generalization capability. In contrast, SEMBA achieves an AUC of 0.99 in link prediction by focusing solely on modeling link existence; however, it fails to explicitly decouple sign semantics, leading to notably degraded sign prediction performance. By contrast, LSWJP employs dual-channel feature decoupling and multi-task joint optimization, maintaining high link prediction accuracy while substantially improving sign prediction, thereby achieving superior overall balance.

As shown in Table \ref{table_mae}, LSWJP consistently achieves significantly lower MAE and RMSE than all baseline methods across all weighted datasets. This result demonstrates that the proposed components—sign-aware embedding, multi-hop structural balance features, and temporal difference modeling—effectively capture the intrinsic correlations between signs and weights, enabling more accurate numerical prediction.

\subsection{Ablation Study}
To systematically evaluate the contribution of each core component in LSWJP, we construct four ablation variants and conduct comparative experiments on the \text{bitcoinotc} dataset. The variants are defined as follows:
\begin{itemize}
    \item \textbf{No-Transformer}: Removes the Transformer-based temporal encoder and directly predicts $G_{t+1}$ using only the last snapshot $G_t$, to verify the necessity of temporal modeling.
    
    \item \textbf{No-embedding}: Disables the sign-aware weighted random walk–generated node embedding $\mathbf{X}_{e,t}^*$, retaining only the structural features $\mathcal{F}_t$ and $\boldsymbol{\Delta}_t$ as input to the encoder.
    
    \item \textbf{No-feature}: Replaces the multi-hop structural balance and temporal difference features with simple sums of positive and negative edge weights as node sign features.
    
    \item \textbf{No-decoupling}: Removes the dual-channel decoupling design and feeds the raw node representation $\mathbf{X}_t$ directly into the Transformer for joint prediction of all three tasks.
\end{itemize}

The experimental results are shown in Table \ref{ablation}, from which we observe the following:

\begin{table}[t]
	\centering
	\caption{Ablation study comparison}
	\begin{tabular}{ccccc}
		\toprule
		Variant        & AUC(L) & AUC(S) & MAE    & RMSE   \\
        \midrule
        No-Transformer & 0.9719 & 0.7408 & 0.1098 & 0.1750 \\
        No-embedding   & 0.9579 & 0.8024 & 0.1016 & 0.1645 \\
        No-feature     & 0.9716 & 0.7745 & 0.1342 & 0.1880 \\
        No-decoupling  & 0.9729 & 0.7856 & 0.1103 & 0.1709 \\
        LSWJP          & 0.9768 & 0.8101 & 0.1000 & 0.1631 \\
		\bottomrule
	\end{tabular}
	\label{ablation}
\end{table}

\begin{itemize}
    \item \textbf{No-Transformer}: Removing the Transformer causes a significant drop of 0.07 in sign prediction AUC, indicating that dynamic evolution patterns are crucial for determining edge polarity. In contrast, link prediction degrades only slightly, suggesting that static snapshots already encode most linkage information.
    
    \item \textbf{No-feature}: Performance degrades substantially across all metrics, especially in weight prediction where MAE increases by 0.034. This demonstrates that simple sums of positive/negative edge weights fail to capture high-order structural balance and temporal dynamics, thereby validating the effectiveness of our multi-hop structural balance and temporal difference mechanisms.
    
    \item \textbf{No-embedding}: Link prediction AUC drops by 0.02, while sign prediction remains relatively stable. This suggests that sign-aware embeddings primarily encode local topological structure, with particular importance for link prediction modeling.
    
    \item \textbf{No-decoupling}: Although link prediction performance remains high, sign prediction AUC decreases by 0.025, indicating that mixed input representations introduce task interference. The dual-channel design effectively enables functional decoupling and collaborative optimization.
\end{itemize}

In summary, removing any component leads to performance degradation in at least one task, and the full LSWJP model achieves the best results across all metrics. This fully validates the necessity and complementarity of each module in the proposed architecture.

\section{Conclusion}\label{sec:conclusion}
This paper proposed a novel joint prediction method, LSWJP, for dynamic signed weighted networks, aiming to achieve the joint modeling of link existence, sign polarity, and edge weight prediction, while maintaining generalization ability to unsigned or unweighted networks. The method first discretized continuous time into an ordered sequence of snapshots. Then, it generated decoupled positive and negative node embeddings via a sign-aware weighted random walk, and designed multi-hop structural balance features and temporal difference features to explicitly capture local balance, high-order topological evolution, and dynamic interaction trends. In terms of model design, a Transformer encoder was adopted to learn the spatial-temporal contextual representations of nodes. A dual-channel decoupling and multi-task joint optimization mechanism was utilized to collaboratively improve the performance of the three prediction tasks. Extensive experiments demonstrated that LSWJP significantly outperformed existing baseline methods on multiple real-world dynamic network datasets. Ablation studies further validated the effectiveness and complementarity of each component.

\bibliographystyle{IEEEtran}  
\bibliography{MyRefs}

\end{document}